\documentclass[12pt]{article}
\usepackage{amssymb}
\usepackage{epsfig}
\begin{document}
%
\setlength{\baselineskip}{0.65cm}
\setlength{\parskip}{0.35cm}
%
\begin{titlepage}
%
\begin{flushright}
CPT-2002/PE.4348 \\
BNL-NT-02/6 \\
RBRC-247 \\
April 2002
\end{flushright}

\vspace*{1.1cm}
\begin{center}
\LARGE
{\bf {Accessing Transversity in Double-Spin}}

\medskip
{\bf {Asymmetries at the BNL-RHIC}}

\vspace*{1.5cm}
\large 
{J.\ Soffer$^{a}$, M.\ Stratmann$^{b}$, and W.\ Vogelsang$^{c,d}$}

\vspace*{1.0cm}
\normalsize
{\em $^a$Centre de Physique Th{\'e}orique CNRS Luminy Case 907,\\
F-13288 Marseille Cedex 09, France}\\

\vspace*{0.5cm}
{\em $^b$Institut f{\"u}r Theoretische Physik, Universit{\"a}t Regensburg,\\
D-93040 Regensburg, Germany}\\

\vspace*{0.5cm}
{\em $^c$Physics Department, Brookhaven National Laboratory,\\
Upton, New York 11973, U.S.A.}\\

\vspace*{0.5cm}
{\em $^d$RIKEN-BNL Research Center, Bldg. 510a, Brookhaven 
National Laboratory, \\
Upton, New York 11973 -- 5000, U.S.A.}
\end{center}

\vspace*{1.5cm}
\begin{abstract}
\noindent
We give upper bounds for transverse double-spin asymmetries in
polarized proton-proton collisions by saturating the positivity
constraint for the transversity densities at a low hadronic 
resolution scale. We consider prompt photon, jet, pion,
and heavy flavor production at the BNL Relativistic Heavy Ion
Collider (RHIC). Estimates of the expected statistical accuracy
for such measurements are presented, taking into account the 
acceptance of the RHIC detectors.
\end{abstract}
\end{titlepage}
\newpage
%
%
\noindent
The partonic structure of spin-1/2 targets at the leading-twist level
is characterized entirely by the unpolarized, longitudinally polarized, 
and transversely polarized distribution functions $f$, $\Delta f$, 
and $\delta f$, respectively \cite{ref:jaffeji}. 
By virtue of the factorization theorem, these non-perturbative
parton densities can be probed universally in a multitude of 
scattering processes as long as some arbitrary hard scale $\mu_F$ can be 
introduced 
to separate long- and short-distance physics. The latter, hard scattering
coefficient functions as well as the $\mu_F$ dependence of the
parton densities can be calculated perturbatively in QCD and confronted
with experiment. By now measurements of helicity averaged deep-inelastic
lepton-nucleon scattering (DIS) have reached a precision that 
theoretical uncertainties become the limiting factor in the extraction of 
unpolarized parton distributions $f$.
Combined experimental and theoretical efforts have led also to an improved
understanding of the spin structure of longitudinally polarized nucleons, 
$\Delta f$, in the past few years. 
Yet many interesting questions still remain unanswered here. 
In particular, nothing is known about the helicity dependent gluon density, 
$\Delta g$. By far the most elusive quantity is, however, the completely 
unmeasured parton content $\delta f$ of transversely polarized nucleons.

Current and future experiments are designed to further unravel the
spin structure of {\em both} longitudinally {\em and} transversely 
polarized nucleons.
Information will soon be gathered for the first time from 
polarized proton-proton collisions at the BNL Relativistic Heavy
Ion Collider (RHIC) \cite{ref:rhic}. The main thrust of the RHIC spin 
program is to hunt
down $\Delta g$ by measuring double-spin asymmetries $A_{\mathrm{LL}}$ 
for various processes in longitudinally polarized $pp$ collisions 
at high energies \cite{ref:rhic}. 
However, collisions of transversely polarized protons will be studied
as well, and the potential of RHIC in accessing transversity 
$\delta f$ in transverse double-spin asymmetries $A_{\mathrm{TT}}$ will be 
examined in this paper for all conceivable scattering processes.
First we briefly review the main features of transversity \cite{ref:ratcliffe}
to elucidate the difficulties in pinning it down experimentally. Next we 
present results
for $A_{\mathrm{TT}}$ for prompt photon and jet production and
comment on pion pair and heavy quark final states. 
Special emphasis is put on realistic estimates of the
statistical precision for such measurements at RHIC including limitations from
the detectors.  We also briefly comment on the Drell-Yan process and 
alternative
methods of filtering for transversity which involve special types of 
fragmentation
functions. It should be stressed that studies of $A_{\mathrm{TT}}$ have been 
presented in the past~[4-7]. These
have usually \cite{ref:ji,ref:jaffesaito} been based 
on the model assumption $\delta f(x,\mu)=\Delta f(x,\mu)$ 
for {\em all} scales $\mu$, which cannot be valid in QCD 
due to the different $\mu$-evolutions
of $\delta f$ and $\Delta f$. Of course, since nothing is known experimentally 
about $\delta f$ one has to rely on some model to study spin asymmetries 
$A_{\mathrm{TT}}$. 
As will be explained below, we make use of the Soffer inequality 
\cite{ref:soffer} to give {\em upper bounds} for $A_{\mathrm{TT}}$. 
This should provide some guidance on
what processes one should focus on in first experimental studies of 
transversity at RHIC. On the other hand,
measurements of $A_{\mathrm{TT}}$ in excess of these bounds
may indicate a new ``spin surprise''. Theoretical uncertainties in
$A_{\mathrm{TT}}$ due to variations of the arbitrary factorization scale
$\mu_F$ are addressed.
In view of all these points and upcoming exploratory studies with transversely 
polarized protons at RHIC, we believe our paper to be timely and useful.

%
The transversity density $\delta f$ is defined
\cite{ref:jaffeji,ref:artru,ref:ralston}, in complete analogy to its 
longitudinally
polarized counterpart $\Delta f$, as the difference of finding a parton 
of flavor $f$ at a scale $\mu$ with momentum fraction $x$ and its spin aligned 
($\uparrow\uparrow$) and anti-aligned ($\downarrow\uparrow$)
to that of the transversely polarized nucleon, i.e.,
\begin{equation}
\label{eq:pdf}
\delta f(x,\mu) \equiv f_{\uparrow\uparrow}(x,\mu) -
                       f_{\downarrow\uparrow}(x,\mu)
\end{equation}
(an arrow always denotes transverse polarization in the following).
The unpolarized densities are recovered by taking the sum in 
Eq.~(\ref{eq:pdf}).
Upon expressing transversely polarized eigenstates as superpositions of
helicity eigenstates, $\delta f$ reveals its helicity-flip, 
chiral-odd nature \cite{ref:jaffeji,ref:artru}. 
Since all QCD and electroweak interactions preserve helicity/chirality, 
$\delta f$
completely decouples from standard inclusive DIS which explains its 
elusiveness.
In addition, a transversity (helicity-flip) gluon density is forbidden
for spin-1/2 targets as helicity changes by two units cannot 
be absorbed \cite{ref:jaffeji,ref:artru,ref:ji2}.
This complete lack of quark-gluon mixing leads to particularly simple, 
non-singlet type scale $\mu$ evolution equations for $\delta f$ which 
are known up to the next-to-leading order (NLO) of QCD 
\cite{ref:nlokernels,ref:wvkern}. 
These kernels have the striking feature that {\em all} Mellin-$n$
moments, $\int_0^1 x^{n-1} \delta f(x,\mu) dx$, {\em decrease} under 
evolution, i.e., transversity ``evolves away'' at all $x$ 
with increasing $\mu$. In particular,  the nucleon's tensor charge, 
given by~\cite{ref:jaffeji} $\sum_f 
\int_0^1 dx\,[\delta f(x,\mu) -\delta \bar{f}(x,\mu)]$, is not conserved 
under QCD evolution.
It is also apparent that a quark sea polarization $\delta \bar{q}(x,\mu)$, 
where $q=u,\, d,\, s$, is vanishingly small at all scales, {\em if} it is 
not already present at the input scale $\mu_0<\mu$
(beyond the leading order (LO) of QCD a numerically tiny $\delta \bar{q}
(x,\mu)$ is generated dynamically under evolution~[11-13].).
Anyway, since most of the required subprocess cross sections are not known in 
the NLO, we limit ourselves for consistency to $\delta f$'s evolved with LO
kernels \cite{ref:artru,ref:baldr,ref:rederiv} only.

The requirement of helicity conservation in hard scattering processes
implies that chirality has to be flipped twice in order to be sensitive
to transversity. One possibility, which we are going to consider in the
following, is to have two transversely polarized hadrons in the initial state
and to measure double-spin asymmetries 
\begin{equation}
\label{eq:att}
A_{\mathrm{TT}}({\cal{P}}) = \frac{\frac{1}{2}
              \left[d\sigma/d{\cal{P}}(\uparrow\uparrow) - 
              d\sigma/d{\cal{P}}(\uparrow\downarrow)\right]}
                                  {\frac{1}{2}
                \left[d\sigma/d{\cal{P}}(\uparrow\uparrow) + 
               d\sigma/d{\cal{P}}(\uparrow\downarrow)\right]}
\equiv \frac{d\delta\sigma/d{\cal{P}}}{d\sigma/d{\cal{P}}}\;\;.
\end{equation}
Here ${\cal{P}}$ stands for any appropriate set of kinematical variables 
characterizing the observed final state.
An alternative is to have only one transversely polarized initial hadron and
a fragmentation process in the final state that is sensitive to transverse
polarization. Such experiments can be carried out at RHIC \cite{ref:rhic} 
as well as in semi-inclusive DIS at the fixed target experiments 
HERMES \cite{ref:hermes} and COMPASS \cite{ref:compass}.
Several different fragmentation processes have been identified as being 
potentially 
suitable for a transversity measurement \cite{ref:ratcliffe}. 
One possibility is to observe the fragmentation of a transversely polarized 
quark into a transversely polarized $\Lambda$  hyperon  as described by 
a transversity fragmentation function $\delta D_q^{\Lambda}$~\cite{ref:lambda}.
Other promising approaches have emerged from considering an asymmetry in the
transverse momentum distribution of a hadron in a jet around the jet axis 
(``Collins effect'' \cite{ref:collinsff}), or the interference between 
$s$- and $p$-waves of a two pion system (``interference fragmentation'' 
\cite{ref:interff}). For example, it has been shown~\cite{ref:boer} that 
combinations of Collins fragmentation functions and transversity 
densities may be partly responsible for sizable azimuthal spin asymmetries 
seen in DIS~\cite{ref:hermes1}, as well as~\cite{ref:ansel} for surprisingly 
large single transverse spin asymmetries discovered in fixed target 
$p^{\uparrow}p\to \pi X$ experiments~\cite{ref:expsingle}.
Attempts are currently being made to study the Collins effect
in $e^+e^-$ scattering, and also to measure the 
interference fragmentation functions there~\cite{ref:belle}.

There are, however, several problems with most such conceivable signatures 
for transversity based on fragmentation spin effects. First of 
all, for all of them the analyzing power is a priori unknown and may 
well be small. Second, in practically all cases, there are 
competing mechanisms for generating the physical observable that
do {\em not} involve transversity.  For example, as
was recently pointed out~\cite{ref:brodsky}, the azimuthal
asymmetries seen in DIS could also result from
``final-state interactions'' of the struck quark, and the large single 
transverse spin asymmetries in $p^{\uparrow}p\to \pi X$ mentioned
above may be explained by non-trivial higher twist effects, described in 
the context of QCD factorization theorems~\cite{ref:stermanqiu}.
It may therefore be difficult to cleanly disentangle transversity from 
other effects. Finally, there is not simply one single ``Collins function'';
rather, it exists for each flavor separately. This increases the 
number of unknowns and may complicate conclusions on more detailed aspects
of transversity~\cite{ref:dboer}. 

For these reasons, we will focus in this paper exclusively on transverse 
double-spin asymmetries of the type~(\ref{eq:att}) at RHIC as 
signatures for transversity. For our observables, the $\delta f$'s we 
are after are the only unknowns in the calculation. In addition, and 
more generally, our study is also motivated by the fact that $pp$ 
collisions at high energies provide a natural environment for determining 
twist-2 parton densities: it is an experience from the unpolarized case 
that in the collider regime theory  calculations based on perturbative 
partonic hard-scattering work best and generally describe experimental 
data well, unlike the fixed-target situation
where sometimes significant discrepancies between data and LO (or NLO) QCD
occur.  It is therefore important to investigate how much $A_{\mathrm{TT}}$ 
measurements at RHIC will tell us about transversity.
 
On the downside, transverse double-spin asymmetries are expected to be 
very small in general \cite{ref:hidaka,ref:ji,ref:jaffesaito}. 
Unpolarized gluons play an important 
or even dominant role in almost all production processes. The complete lack
of gluon induced subprocesses in case of transverse polarization strongly
dilutes $A_{\mathrm{TT}}$.
In addition, $A_{\mathrm{TT}}$ in Eq.~(\ref{eq:att}) is further 
diminished by the requirement of a double chirality flip in the transversely
polarized cross section which excludes some of the 
``standard'' $2\rightarrow 2$ amplitudes to contribute, whereas the 
remaining ones are color-suppressed. 
For instance, for $qq$ scattering only the interference between the
two LO Feynman diagrams has a quark line connecting
the two polarized incoming hadrons as necessary for a chirality flip. 
It should be also noted that although $\delta f$ 
is the only unknown in an analysis of $A_{\mathrm{TT}}$, it always appears
``quadratic'' in doubly polarized scattering. In general the task is to find
for each process a kinematical region where gluons contribute as little
as possible to the unpolarized cross section in Eq.~(\ref{eq:att}). 
This is expected for the production of a final state with high transverse 
momentum $p_T$, e.g., a jet or prompt photon.
On the other hand, for large $p_T$ the cross section becomes small
which reduces the statistical precision,
$\delta A_{\mathrm{TT}}$ with which $A_{\mathrm{TT}}$ can be
measured. Also, as mentioned above, $\delta f$ evolves away if probed at 
large hard scales
$\mu_F\simeq p_T$. Thus one always has to find the best compromise 
between the size of $A_{\mathrm{TT}}$ on the one hand, and 
$\delta A_{\mathrm{TT}}$ on the other. In any case, for RHIC $A_{\mathrm{TT}}$
should be larger than about 0.001 to be detectable, since 
systematic uncertainties from relative polarization measurements 
are presently at best expected to be of that order \cite{ref:blandbunce}.

In principle the most favorable reaction for determining
transversity is the Drell-Yan process, $pp\to \mu^+\mu^-$, 
which proceeds exclusively through $q\bar{q}$ annihilation in LO 
without any gluonic contribution.
Not surprisingly, transversity was first studied theoretically
in the context of the Drell-Yan process \cite{ref:ralston}, 
and several phenomenological studies have been performed since
\cite{ref:dy2,ref:dy1}. However, a recent NLO study 
of upper bounds for $A_{\mathrm{TT}}$ \cite{ref:dynlo}, 
using the same model for $\delta f$ as described below,
has revealed that the limited muon acceptance for the RHIC
experiments \cite{ref:rhic} threatens to make a measurement of
transversity in this channel elusive. In particular, the dependence of 
$A_{\mathrm{TT}}$
on the rapidity $y$ of the muon pair, which would be sensitive to the
shape of $\delta f$, receives a substantial 
relative statistical error \cite{ref:dynlo}.
It should be also kept in mind that only the product of quark and
anti-quark transversity densities is probed, and in $pp$ scattering
the latter may well be much smaller than assumed in the model below,
due to the lack of the gluon splitting $g\to q\bar{q}$ discussed above.
In the following we demonstrate that, although $A_{\mathrm{TT}}$ is rather
minuscule, jet and prompt photon production
can be nevertheless a useful tool to decipher transversity at RHIC.

To be specific, according to the factorization theorem the fully
differential transversely polarized cross section $d\delta\sigma$ on the
right-hand-side of Eq.~(\ref{eq:att}) for, say, the production of 
two massless partons $c$ and $d$ with transverse momentum $p_T$,
azimuthal angle $\Phi$ with respect to the initial spin axis, 
and pseudorapidities $\eta_c$ and $\eta_d$ reads 
\begin{equation} 
\label{eq:xsec}
\frac{d^4\delta \sigma}{dp_T d\eta_c d\eta_d d\Phi} =
2p_T\, \sum_{a,b}\, 
x_a \delta f_a (x_a,\mu_F)\,
x_b \delta f_b (x_b,\mu_F)\;  
\frac{d^2\delta\hat{\sigma}}{dt d\Phi}(s,t,u,\Phi, \mu_F,\mu_R)\;\; ,
\end{equation}
where $d\delta\hat{\sigma}$ is the partonic cross section. 
In the partonic center of mass system (c.m.s.), the
Mandelstam variables are given by
\begin{equation}
\label{eq:stu}
s =  x_a x_b S\;,\;
t = -x_a \sqrt{S} p_T e^{-\eta_c}\;,\;
u = -x_a \sqrt{S} p_T e^{-\eta_d}\;
\end{equation}
where $\sqrt{S}$ denotes the available hadronic c.m.s.\ energy and 
\begin{equation}
\label{eq:xaxb}
x_a = \frac{2p_T}{\sqrt{S}} \cosh\left[\frac{1}{2}(\eta_c-\eta_d)\right]
      e^{(\eta_c+\eta_d)/2}\;,\;
x_b = \frac{2p_T}{\sqrt{S}} \cosh\left[\frac{1}{2}(\eta_c-\eta_d)\right]
      e^{-(\eta_c+\eta_d)/2}\;\;.
\end{equation}
Since discrepant results for the partonic cross sections
$d\delta\hat{\sigma}$ can be found
in the literature~[4-7],
we have recalculated and listed them in Table~\ref{tab:xsecs}. 
We fully agree with the expressions given in Tab.~II of \cite{ref:jaffesaito},
except for an overall factor $2\pi$ in the normalization of the cross
sections in that paper (such a factor does not affect spin asymmetries).  
For completeness we give here also the result for heavy flavor production
which will be briefly discussed below.
The entries for $\delta |M|^2$ in Tab.~\ref{tab:xsecs} have
to be multiplied by an appropriate factor to yield the 
partonic cross sections relevant for Eq.~(\ref{eq:xsec}):
\begin{equation}
\label{eq:prefactor}
\frac{d^2\delta\hat{\sigma}}{dt d\Phi}= \frac{2\alpha_s^2(\mu_R)}{9s^2}\,
\frac{t\,u}{s^2}\,\cos(2\Phi)\,\delta |M|^2\;\;, 
\end{equation}
with $\mu_R$ denoting the renormalization scale.
In the case of prompt photon production, $\alpha_s^2$ in
Eq.~(\ref{eq:prefactor}) has to be replaced by $\alpha_s \alpha e_q^2$ with
$e_q$ the electrical charge of the incoming quark $q$. For production
of heavy flavors via $q\bar{q}\to Q\bar{Q}$, 
the factor $tu/s^2$ in Eq.~(\ref{eq:prefactor}) becomes
$(t'u'-m^2 s)/s^2$, where $m$ is the mass of the heavy quark, and 
the reduced Mandelstam variables $t',u'$ are
given as $t'=t-m^2$, $u'=u-m^2$. Expressions~(\ref{eq:stu}) and~(\ref{eq:xaxb})
remain valid with $t\to t',\,u\to u'$ and $p_T\to m_T=\sqrt{p_T^2+m^2},\,
\eta_{c,d}\to y_{c,d}$ with the heavy flavor rapidities $y_{c,d}$.
%
%
\begin{table}[h,t]
\renewcommand{\arraystretch}{1.2}
\begin{center}
\begin{tabular}{|c|c|c|}
\hline
final state & subprocess & $\delta |M|^2$ \\ \hline \hline
 & $qq\to qq$ & $2 s^2/(3 t u)$ \\
 & $qq^\prime \to qq^\prime$ & -- \\
jets and & $q\bar{q} \to q\bar{q}$ & $2 - 2 s/(3 t)$ \\ 
inclusive hadrons & $q\bar{q} \to q^\prime \bar{q}^\prime$ & $2$ \\
 & $q\bar{q}^\prime \to q \bar{q}^\prime$ & -- \\
 & $q\bar{q} \to gg$ & $16 s^2/(3 t u)-12$ \\  \hline \hline
prompt photons & $q\bar{q} \to g\gamma$ & $4 s^2/(t u)$ \\ \hline \hline
heavy quarks  & $q\bar{q} \to Q \bar{Q}$ & $2$ \\ 
\hline
\end{tabular}
\end{center}
\caption{\label{tab:xsecs}\sf LO $2\rightarrow 2$ squared matrix elements 
contributing in the case of transverse polarization. Note that each entry 
$\delta |M|^2$ has to be multiplied by a prefactor, 
see Eq.~(\ref{eq:prefactor}),
to obtain the actual partonic cross section $d\delta\hat{\sigma}$ relevant for
Eq.~(\ref{eq:xsec}).}
\end{table}

While the corresponding unpolarized cross section 
$d^4\sigma/dp_T d\eta_c d\eta_d d\Phi$ is $\Phi$-independent and
can be trivially integrated over $\Phi$ yielding the usual factor 
$2\pi$, the factor $[t\,u\,\cos(2\Phi)/s^2]$ in Eq.~(\ref{eq:prefactor}) is
a very characteristic feature of double transverse polarization of the 
incoming partons\footnote{For arbitrary azimuthal angles $\Phi_{a,b}$ of the 
initial transverse spin vectors the cross section actually has 
the angular dependence $\cos(2\Phi-\Phi_a-\Phi_b)$~\cite{ref:ralston}.}. 
It would obviously integrate to zero, and therefore each 
quadrant in $\Phi$ should be added with the 
appropriate sign upon integration~\cite{ref:cpr}, 
i.e., $\left(\int_{-\pi/4}^{\pi/4}-
\int_{\pi/4}^{3\pi/4}+\int_{3\pi/4}^{5\pi/4}-\int_{5\pi/4}^{7\pi/4} \right) 
\cos(2\Phi) d\Phi=4$. This way of combining cross sections has
the advantage that it leads to much higher rates as compared to 
considering only~\cite{ref:jaffesaito} a small bin around, say, $\Phi=0$.
An alternative approach would be to integrate the spin-dependent
cross section over $\Phi$ with a weight factor $\cos(2\Phi)$. By 
using other Fourier harmonics as weight factors, one could actually
verify experimentally the $\cos(2\Phi)$ dependence of $A_{\mathrm{TT}}$
predicted by QCD.

%
Before we can perform numerical studies of $A_{\mathrm{TT}}$ we have 
to specify the $\delta f$. Since nothing is known experimentally 
about them one 
has to fully rely on models to study spin asymmetries $A_{\mathrm{TT}}$. 
The only 
guidance so far is provided by the Soffer inequality \cite{ref:soffer}
\begin{equation}
\label{eq:soffer}
2\left|\delta q(x)\right| \leq q(x) + \Delta q(x)
\end{equation}
which gives an upper bound for $\delta f$ in terms
of the unpolarized and helicity-dependent quark distributions 
$q$ and $\Delta q$, respectively. As in \cite{ref:dynlo} we utilize
this inequality by saturating the bound at some low input scale
$\mu_0\simeq 0.6\,\mathrm{GeV}$ using the LO GRV \cite{ref:grv98} 
and GRSV (``standard scenario'') \cite{ref:grsv2000} densities
$q(x,\mu_0)$ and $\Delta q(x,\mu_0)$, respectively. 
For $\mu>\mu_0$ the transversity densities $\delta f(x,\mu)$ are then obtained
by straightforwardly solving the evolution equations using the appropriate 
LO kernels \cite{ref:artru,ref:baldr,ref:rederiv}. It should be noted 
that the inequality (\ref{eq:soffer}) is preserved under
evolution in LO and NLO (${\overline{\rm MS}}$), as was shown 
in~\cite{ref:wvkern,ref:dy2,ref:sofferevol}. Obviously,
the sign to be used when saturating the inequality is at our
disposal. Our calculations below will always refer to choosing
a positive sign throughout.  This is expected to result in the 
largest possible asymmetries, since then all transversity
densities have the same sign at the input scale, and the partonic
cross sections in Table~\ref{tab:xsecs} 
are all positive.  Note that, thanks to the
especially simple evolution of the transversity densities, 
the sign is preserved under $Q^2$-evolution, for each parton
density individually. We have varied the sign in
the saturation of Eq.~(\ref{eq:soffer}) independently for all parton densities;
however the asymmetry always came out smaller than the one obtained
for only positive signs.  It is important, however, to point out
that the actual {\em sign} of the asymmetry cannot be predicted
in this way: for example, we could saturate the inequality
with positive signs for the quark densities and with negative signs
for the antiquarks. Then all $q\bar{q}$ scatterings will give
negative contributions, and only the $qq$ and $\bar{q}\bar{q}$ channels
remain positive.  Therefore, the asymmetry may potentially become
negative in certain kinematical regions, but it will never be as large  
{\it in absolute value} as our default one obtained for choosing only
positive signs in saturating the Soffer inequality. 

Using these densities we can present the maximally possible spin asymmetries 
$A_{\mathrm{TT}}$. It should finally be kept in mind that by saturating 
Eq.~(\ref{eq:soffer}) at a somewhat higher scale, e.g., 
$\mu_0\simeq 1\,\mathrm{GeV}$, one would obtain in general somewhat 
(but not very much) larger $A_{\mathrm{TT}}$'s
as compared to our results in Figs.~\ref{fig:1jet} and \ref{fig:photon}
below. 
It does not seem very natural, however, that Soffer's inequality
is saturated at scales much larger than $\mu_0\simeq 1\,\mathrm{GeV}$. 
Therefore we believe that our results represent fairly realistic
upper bounds for $A_{\mathrm{TT}}$. It is also quite possible that
Eq.~(\ref{eq:soffer}) is actually far from being saturated. 
%
%
\begin{figure}[t]
\begin{center}
\vspace*{-0.8cm}
\epsfxsize=0.7\textwidth
\epsfbox{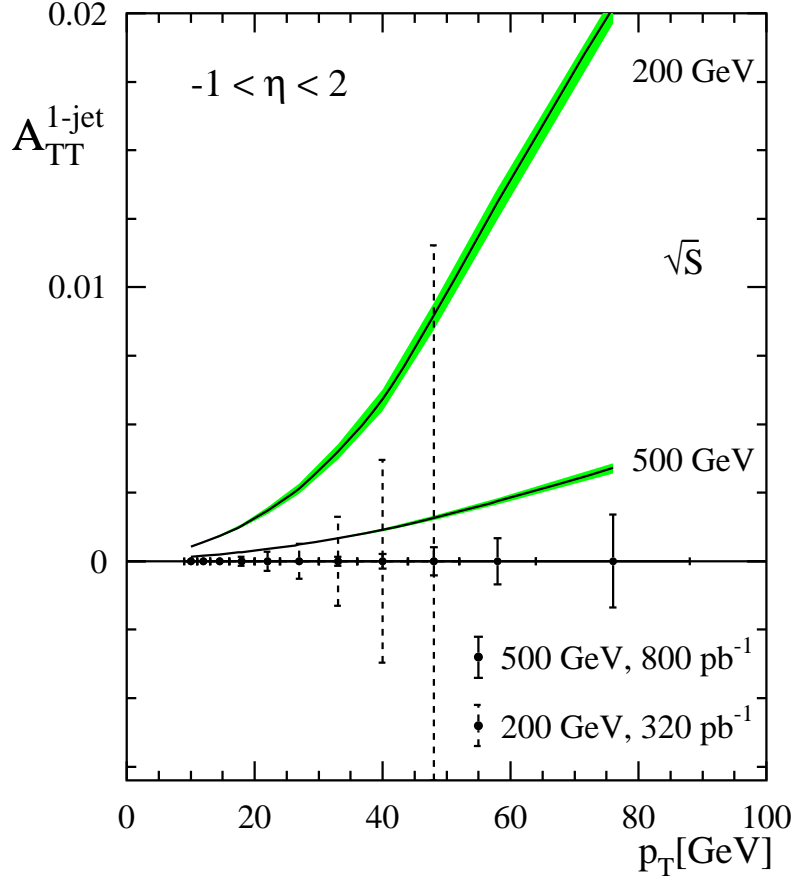}
\caption{\label{fig:1jet}\sf ``Maximally possible'' $A_{\mathrm{TT}}$ for 
single-inclusive jet production at RHIC c.m.s.\ energies 
of $200\,\mathrm{GeV}$ and $500\,\mathrm{GeV}$
as a function of $p_T$. Jet rapidities are integrated
over the detector acceptance ($-1\le \eta \le 2$). 
The shaded bands represent the theoretical uncertainty
in $A_{\mathrm{TT}}$ if $\mu_F$ is varied in the
range $p_T/2\le \mu_F \le 2 p_T$. Also indicated
as ``error bars'' is the expected statistical accuracy
for certain bins in $p_T$.}
\end{center}
\end{figure}
In this case all spin asymmetries would be much smaller and not accessible 
by RHIC at all. However, in non-relativistic quark models one has
$\delta q=\Delta q$ which also leads to sizable transversity densities, 
but may be at variance with Soffer's inequality. 

Let us first turn to single-inclusive jet production
as jets are copiously produced at high energies and are theoretically
rather well understood. The corresponding
double-spin asymmetry $A_{\mathrm{TT}}$ is shown in Fig.~\ref{fig:1jet}
as a function of the jet transverse momentum $p_T$. The rapidities 
in Eq.~(\ref{eq:xsec}) are integrated over the range
$-1\le \eta \le 2$ which is accessible with in the STAR experiment at RHIC.
As expected, $A_{\mathrm{TT}}$ is rather small due to the dominance
of the gluon induced processes in the unpolarized cross section.
In the $p_T$ range shown, the $qg$ subprocess alone always contributes about
$50\%$ of the total rate. At low $p_T$ also the $gg$ subprocess contributes
significantly. In case of transversity, $d\delta \sigma$, 
the $qq\to qq$ (and $\bar{q}\bar{q}\to \bar{q}\bar{q}$) process 
is color-suppressed as $1/N_C$ since only
the interference of different amplitudes has the required
helicity-flip properties, whereas the $q\bar{q}$ channel  suffers 
from the presumed scarcity of antiquarks in the initial state 
for $pp$ collisions. Nevertheless, for our model transversity densities for
which we also saturate the antiquark densities at the initial scale, the 
asymmetry at $\sqrt{S}=500\,\mathrm{GeV}$ is mainly driven by the $q\bar{q}$ 
annihilation reaction unless $p_T$ becomes larger 
then about $40\,\mathrm{GeV}$ 
and the $qq$ channel starts to take over. The sizable antiquark
density we propose would necessarily be of nonperturbative origin.
Its presence could provide information on the breaking of chiral 
symmetry in QCD, as is true for transversity in 
general~\cite{ref:jaffetalk}.

The bands in Fig.~\ref{fig:1jet} indicate the theoretical uncertainty
in $A_{\mathrm{TT}}$ due to variations of the factorization scale 
in the range $p_T/2\le \mu_F \le 2p_T$. We have always identified
the renormalization scale $\mu_R$ with $\mu_F$ since in the LO
of QCD $\alpha_s$ drops out of the asymmetry (\ref{eq:att}) anyway. Also shown 
in Fig.~\ref{fig:1jet} is the expected statistical accuracy for a 
measurement of 
$A_{\mathrm{TT}}$ at c.m.s.\ energies $\sqrt{S}$ of
200 and 500~GeV for certain bins in $p_T$, using an integrated
luminosity of 320 and 800~$\mathrm{pb}^{-1}$, respectively, and
assuming a polarization of $70\%$ for each beam. It turns out that
statistical errors would allow sensible measurements up to a
$p_T$ of about 40 (70) GeV for $\sqrt{S}=200$ (500) GeV.
The region beyond that would only become accessible through a 
substantial upgrade in luminosity.
However, for the smallest $p_T$ bins, $A_{\mathrm{TT}}$ is rather
tiny, in particular for $\sqrt{S}=$500~GeV, and limitations from
the systematical uncertainties become important here,
making a measurement at modest $p_T$ values a very challenging task. 

Instead of detecting a jet one can also consider  identifying a leading 
hadron (pion). Single-inclusive pion production has a smaller
rate though and thus larger statistical errors. From 
a theoretical point of view it is less clean than jet
production as it involves also the modeling of the fragmentation
of a final state parton into the observed pion.
On the other hand, it is an old idea~\cite{ref:fontannaz} that
tagging on {\em two} back-to-back pions could in principle 
help to overcome the difficulties of small transverse spin asymmetries
by reducing the contribution from gluon induced processes in the unpolarized 
cross section: since the fragmentation functions for $g\to \pi^+$ and
$g\to \pi^-$ are the same, the cross section combination~\cite{ref:fontannaz}
$\sigma^{\pi^+\pi^+}-\sigma^{\pi^+\pi^-}-\sigma^{\pi^-\pi^+}+
\sigma^{\pi^-\pi^-}$, where in each case the two pions are in opposite 
azimuthal hemispheres, completely
eliminates the $qg\rightarrow qg$ and $gg\rightarrow gg$ channels 
(although not $gg\rightarrow q\bar{q}$). The corresponding double-spin 
asymmetry,
\begin{equation}
A_{\mathrm{TT}}\equiv 
\frac{\delta\sigma^{\pi^+\pi^+}-\delta\sigma^{\pi^+\pi^-}-
\delta\sigma^{\pi^-\pi^+}+\delta\sigma^{\pi^-\pi^-}}
{\sigma^{\pi^+\pi^+}-\sigma^{\pi^+\pi^-}-\sigma^{\pi^-\pi^+}+
\sigma^{\pi^-\pi^-}} \; ,
\end{equation}
is indeed sizable, reaching easily values as large as $10\%$ in certain
regions of $p_T$ for the pions. However, the catch here is that the 
statistical error $\delta A_{\mathrm{TT}}$ for this combination of cross 
sections
is not only, as usually the case, proportional to the inverse of
the square root of the counting rate, 
\begin{equation}
\delta A_{\mathrm{TT}} \propto \left(\sigma^{\pi^+ \pi^+}  -
\sigma^{\pi^+ \pi^-}-\sigma^{\pi^- \pi^+} + \sigma^{\pi^- \pi^-}
\right)^{-1/2} \; ,
\end{equation}
but rather to
\begin{equation}
\delta A_{\mathrm{TT}} \propto 
\left(\sigma^{\pi^+ \pi^+}  -
\sigma^{\pi^+ \pi^-}-\sigma^{\pi^- \pi^+} + \sigma^{\pi^- \pi^-}
\right)^{-1/2} \;
\sqrt{\frac{\sigma^{\pi^+ \pi^+} 
+ \sigma^{\pi^+ \pi^-} + \sigma^{\pi^- \pi^+}+ \sigma^{\pi^- \pi^-} }
{\sigma^{\pi^+ \pi^+}  -
\sigma^{\pi^+ \pi^-}-\sigma^{\pi^- \pi^+} + 
\sigma^{\pi^- \pi^-}}} \; .
\end{equation}
The extra factor always renders $|\delta A_{\mathrm{TT}}|$ much bigger than
$|A_{\mathrm{TT}}|$ itself.

In Fig.~\ref{fig:photon} we show our ``maximal prediction'' for
$A_{\mathrm{TT}}$ for prompt photon production. At lowest order, 
the basic calculation is similar to that for jet production. Our
predictions are made for $|\eta |\le 0.35$ as relevant for
the PHENIX experiment. We take into account that only half the
azimuth is covered by the detectors in this experiment.  We also
assume that in experiment an isolation criterion is imposed on the
photon, meaning that events are vetoed if much hadronic
energy is found in the vicinity of the photon. We adopt the 
isolation criterion proposed in~\cite{ref:frixione}, which 
eliminates an (unwanted) contribution to the cross section
related to photons produced in jet fragmentation. 
The resulting transverse-spin asymmetries for photons are larger 
than for jets; however, unfortunately the same is true for the expected 
statistical errors. Nevertheless, for transverse momenta of the 
photon not very  much larger than 10 GeV also this measurement may 
provide useful information about the transversity densities.
%
%
\begin{figure}[th]
\begin{center}
\vspace*{-0.8cm}
\epsfxsize=0.7\textwidth
\epsfbox{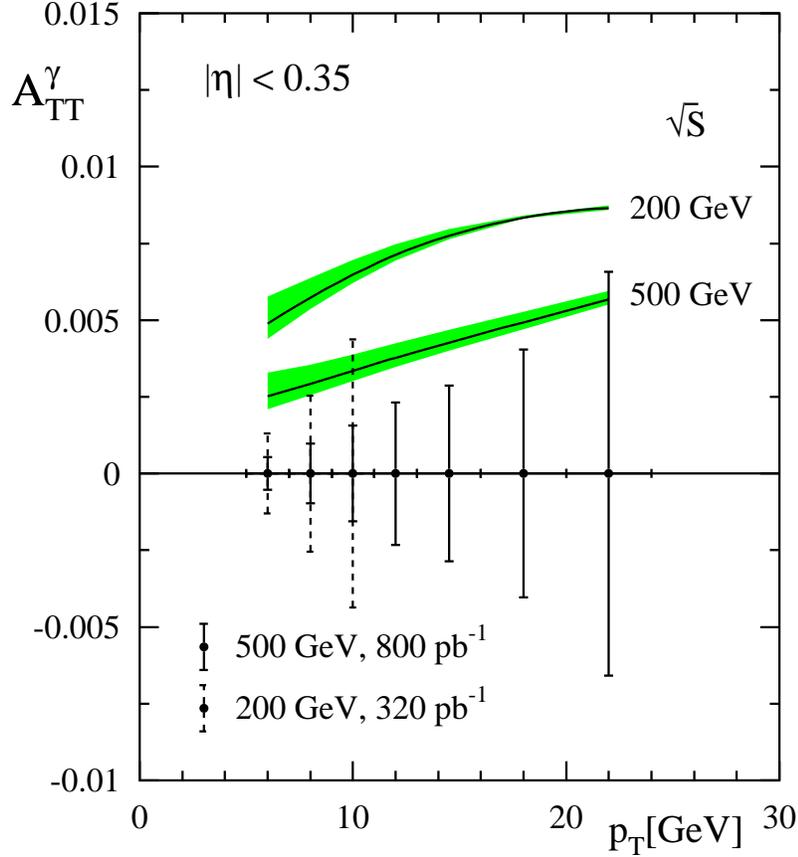}
\caption{\label{fig:photon}\sf Same as in 
Fig.~\ref{fig:1jet} but now for prompt photon production.
The photon rapidity has been integrated over the range
$|\eta |\le 0.35$ accessible at RHIC, and only half of the
full azimuth is taken into account.}
\end{center}
\end{figure}

We note that heavy flavor production turns out to be not particularly useful 
due to the strong dominance of the gluon-gluon fusion process in the 
unpolarized rate in all relevant kinematical regions.

%
In conclusion, accessing transversity at RHIC 
through leading-twist double-spin asymmetries in ``standard''
processes like  jet or prompt photon production appears to be
a difficult, albeit not completely impossible, task. 
In general, even the ``maximally allowed'' transverse double-spin 
asymmetries $A_{\mathrm{TT}}$ are very small, requiring great experimental 
efforts to detect them. Since jets (and photons) are produced very
copiously at RHIC, the main limitation may often be the systematical
rather than the statistical error for measurements
of $A_{\mathrm{TT}}$. Nonetheless, if the $\delta f$, $\delta \bar{f}$ 
arise through evolution from an input at some low scale that saturates 
(or nearly saturates) Soffer's inequality, RHIC will see signatures of 
transversity in these channels, complementing information on transversity 
gained from other sources at RHIC and elsewhere. 

We finally emphasize -- in the spirit of Ref.~\cite{ref:jaffesaito} --
that the (expected) general smallness of $A_{\mathrm{TT}}$, perhaps best
summarized~\cite{ref:jaffesaito} by $|A_{\mathrm{TT}}|\ll |A_{\mathrm{LL}}|$, 
where $A_{\mathrm{LL}}$ is the longitudinal counterpart of $A_{\mathrm{TT}}$ 
for a certain reaction, is also to some extent a virtue: it relies on 
Soffer's inequality and on other important aspects of transverse-spin in QCD, 
such as the peculiar evolution of transversity (non-singlet type, strong 
suppression at  small-$x$) and the kinematical and color-suppression of the
hard scatterings of transversely polarized partons. The experimental finding
of much larger $A_{\mathrm{TT}}$ than the ones presented here would 
constitute a spin puzzle of proportions similar to the ``spin crisis'' 
of the late eighties in the longitudinally polarized case.

%
We are grateful to R.L.\ Jaffe and M.\ Grosse Perdekamp for helpful comments.
M.S.\ thanks A.\ Sch\"{a}fer for discussions and 
RIKEN and Brookhaven National Laboratory for hospitality and support. 
W.V.\ is grateful to RIKEN, Brookhaven National Laboratory and the U.S.
Department of Energy (contract number DE-AC02-98CH10886) for
providing the facilities essential for the completion of this work.
The work of M.S.\ was supported by DFG and BMBF.
\newpage

%
\end{document}